\documentclass[12pt]{article}
\usepackage[T1]{fontenc}

\usepackage[left=1in, right=1in, top=1in, bottom=1in]{geometry}

\usepackage[singlespacing]{setspace}

\usepackage{authblk, amsmath, amssymb, amsfonts, mathptmx, multirow, graphicx, paralist, rotating, siunitx, hyperref}

\usepackage{natbib}
\setcitestyle{aysep={}}

\renewcommand\Affilfont{\normalfont\small}
\makeatletter
\renewcommand\AB@affilsepx{, \protect\Affilfont}
\makeatother

\title{Bivariate Analysis of Birth Weight and Gestational Age Depending on Environmental Exposures: \\Bayesian Distributional Regression with Copulas}

\author[1]{Jonathan~Rathjens}
\author[2]{Arthur~Kolbe}
\author[2]{J\"urgen~H\"olzer}
\author[1]{Katja~Ickstadt}
\author[3]{Nadja~Klein}

\affil[1]{Technische Universit\"at Dortmund}
\affil[2]{Ruhr-Universit\"at Bochum}
\affil[3]{Humboldt-Universit\"at zu Berlin}

\begin{document}

\maketitle

\begin{abstract}
In this article, we analyze perinatal data with birth weight (BW) as primarily interesting response variable. Gestational age (GA) is usually an important covariate and included in polynomial form. However, in opposition to this univariate regression, bivariate modeling of BW and GA is recommended to distinguish effects on each, on both, and between them. Rather than a parametric bivariate distribution, we apply conditional copula regression, where marginal distributions of BW and GA (not necessarily of the same form) can be estimated independently, and where the dependence structure is modeled conditional on the covariates separately from these marginals. In the resulting distributional regression models, all parameters of the two marginals and the copula parameter are observation-specific. 
Besides biometric and obstetric information, data on drinking water contamination and maternal smoking are included as environmental covariates. While the Gaussian distribution is suitable for BW, the skewed GA data are better modeled by the three-parametric Dagum distribution. The Clayton copula performs better than the Gumbel and the symmetric Gaussian copula, indicating lower tail dependence (stronger dependence when both variables are low), although this non-linear dependence between BW and GA is surprisingly weak and only influenced by Cesarean section. A non-linear trend of BW on GA is detected by a classical univariate model that is polynomial with respect to the effect of GA. Linear effects on BW mean are similar in both models, while our distributional copula regression also reveals covariates' effects on all other parameters. 
\end{abstract}


\section{Introduction}\label{sec:intro}

When analyzing perinatal (newborn infants') data with birth weight (BW) as the response variable of primary interest, it is essential to adjust for gestational age (GA, duration of pregnancy), which is often reported as the quantitatively most important covariate (e.g., \citealp{Skj00, Fan07, Wei14, Fre08}). Augmenting linear models, it may be included  as a polynomial or in other parametric functional forms (e.g., \citealp{Gar95, Sal07}). Another widespread alternative is a binary response with a class such as `small for GA' (e.g., \citealp{Gag03, Pol09, Tho01}). 

In contrast to univariate approaches like these, consideration of bivariate (or multivariate) outcomes is frequent in biometric research, such as meta-analysis \citep{Ber98}, clinical trials \citep{Bra02}, dose-response-modeling in developmental toxicology \citep{Reg99}, measuring heavy metal concentrations in the environment \citep{Poz19}, or simultaneous estimation of human exposure to jointly occurring  chemicals \citep{Sy20}. In gynecological and obstetric research, modeling of a bivariate response comprising of both BW and GA is recommended (e.g., \citealp{Gag03, Fan07, Sch10}), but not common. 

When following this recommendation, several  aspects have to be considered. Apart from the possible influences of covariates on both BW and GA, the dependence between these two outcomes is still very relevant and potentially non-linear. Additionally, the form and degree of this dependence may vary with the covariates, e.g., when taking Cesarean sections into account. A convenient way to address all is to apply a copula model \citep{Nel06} for the joint distribution of BW and GA, such that their univariate marginal distributions and their dependence structure are considered and all distribution parameters (not only the means) estimated depending on covariates (distributional copula regression). This approach is more flexible than a classical bivariate regression model with one correlation parameter. 

There is a vast literature on copula modeling, also in the regression context. For instance,  simultaneous maximum likelihood estimation (MLE) in vector generalized linear models have been developed \citep{Son09} using Gaussian copulas with correlation matrices not depending on covariates.   Modeling of covariates' effects on the parameter of various copula families has been worked on \citep{Aca13} as well as  MLE in regression with several outcomes joint by copulas \citep{Kra13}.  Bayesian copula application is also found in human biology \citep{Val18}. Another approach for regression problems is to represent the multivariate density by a (D-)vine copula \citep{Kra17, Coo20}. 

More flexible non-Gaussian copula families assume less symmetries for the data; instead, upper or lower tail dependence can be modeled. As for the marginal distributions, the possibility of non-Gaussian modeling is  desirable to handle asymmetric data as found for BW and GA. Additionally, other parameters than the mean are of interest and potentially dependent on covariates; e.g., the standard deviation of BW may vary between the sexes.  Altogether, such copula models are recommendable for many data situations with asymmetry and unknown or unusual dependence structures. Copulas are widely applicable to bivariate or multivariate biometric analyses in medicine and life sciences, such as those named above.

The estimation of all the parameters in this complex framework is possible with a Bayesian approach to distributional copula regression \citep{Kle16}: Besides modeling the two response variables depending on several covariates, these models comprise of another semiparametric predictor to independently model the parameter of one-parametric copulas between the two of them in a flexible manner. The full procedure is outlined in Figure~\ref{fig:flow}. It is a natural approach to analyze a bivariate response with an asymmetric joint distribution and skewed marginals as found in our data (Figure~\ref{fig:resp}). In the same context, bivariate copula regression with GA and low BW measured as a continuous and a binary variable, respectively, have been considered \citep{Kle19} conditional on various biometric and clinical variables in a spatial context. We now apply such a model with two continuous marginal distributions.

A great advantage of the Bayesian treatment based on Markov chain Monte Carlo (MCMC) simulations is the direct availability of uncertainty estimates via Bayesian credible intervals. Such measures contribute to a better understanding and work well also in small data sizes. 

We compare the copula model to a standard univariate approach, a regression of BW depending on GA modeled as a polynomial. To this end, the distribution of BW conditional on GA obtained from the bivariate copula is numerically evaluated using random numbers drawn from it; thus, we preserve more information from the joint analysis than just the marginal. 

On the data side, our study is based on perinatal registry data from North Rhine-Westphalia (NRW), Germany, which contain many biometric and medical variables on mother, child and birth as well as further possibly influencing factors such as maternal smoking. 

Within our larger `\textbf{PerSpat}' (Perfluoroalkyl Spatial) project, we conduct a study on the general population living in NRW, to evaluate associations of Perfluorooctanoic Acid (PFOA) and other perfluorinated compounds with perinatal parameters, especially the newborn children's weight. PFOA concentrations in some of the region's drinking water resources have been confirmed to be very high, following contamination incidents prior to 2006 \citep{Sku06}. PFOA is suspected to have adverse health affects with regard to fetal development, among others \citep{Joh14, Lam14}. We have analyzed NRW's available PFOA concentration data, measured in drinking water supply stations and networks, in order to assess the residents' external exposure from Summer 2006 on \citep{Rat20}. We now spatio-temporally assign these information on local PFOA concentrations in drinking water to our birth data. 

The article is structured as follows. In Section~\ref{sec:project}, we specify our research questions and present the data in more detail. The applied bivariate copula families and Bayesian distributional regression are outlined in Section~\ref{sec:cop}. Identification of the best model within this class, also with respect to marginal distributions, is reported in Section~\ref{sec:choice}. Substantive analyses follow in Section~\ref{sec:res}, with interpretation of the bivariate copula regression results (Section~\ref{sec:eval}), identification and evaluation of the univariate polynomial model with a comparison to the copula model (Section~\ref{sec:comp}), as well as integration of local PFOA concentrations to our models (Section~\ref{sec:expos}). Section~\ref{sec:disc} discusses some modeling aspects, substantive conclusions from both models and future perspectives for further data analyses.

\section{Research Questions and Data Situation}\label{sec:project}

We here focus on a region along the upper course of the river Ruhr in NRW, precisely the town of Arnsberg: It is of particular interest due to variable and in parts very high PFOA concentrations in drinking water during several years from about 2004 on. Furthermore, the external exposure to PFOA is well assessed and confirmed by human biomonitoring data on the internal exposure of a cohort of residents \citep{Hoe08}. A constrained data analysis also eases computability. 

We investigate, which family of one-parametric copulas, which families of marginal distributions and which linear predictors are most suitable for the given perinatal registry data. Using the identified copula model, we compare the results to a standard univariate regression model for the birth weight (BW) and estimate the effects of biometric, perinatal, environmental and socio-economic covariates on BW, gestational age (GA) and their dependence.

The perinatal registry data are collected by all hospitals and are combined and processed by the quality assurance office residing with the state medical association, for the purpose of quality assurance in obstetrical health care. Within our larger \textbf{PerSpat} project, we use these secondary data from 2003 until 2014, comprising of about 1.7 million records and more than 200 biometric, medical and social variables on mother and child, pregnancy, birth and treatment.  They are anonymized by removing all personal information apart from the postal code of the mother's residence. Further data cleansing steps are performed, in particular regarding the plausibility of GA. Analyses are restricted to singleton births, to children born alive without malformations and to postal codes within NRW. When restricted to the town of Arnsberg, we observe 6442 birth cases within the period from 2003 until 2014. We remove those where values of relevant variables are not given, leaving a total of 4451 observations.

\begin{figure}[tp]
\caption{Observations of birth weight and gestational age (density of the point cloud represented by shading: darker shade is for more points, + stands for a single isolated point).}\label{fig:resp}
\begin{center}
	\includegraphics[width=.49\textwidth]{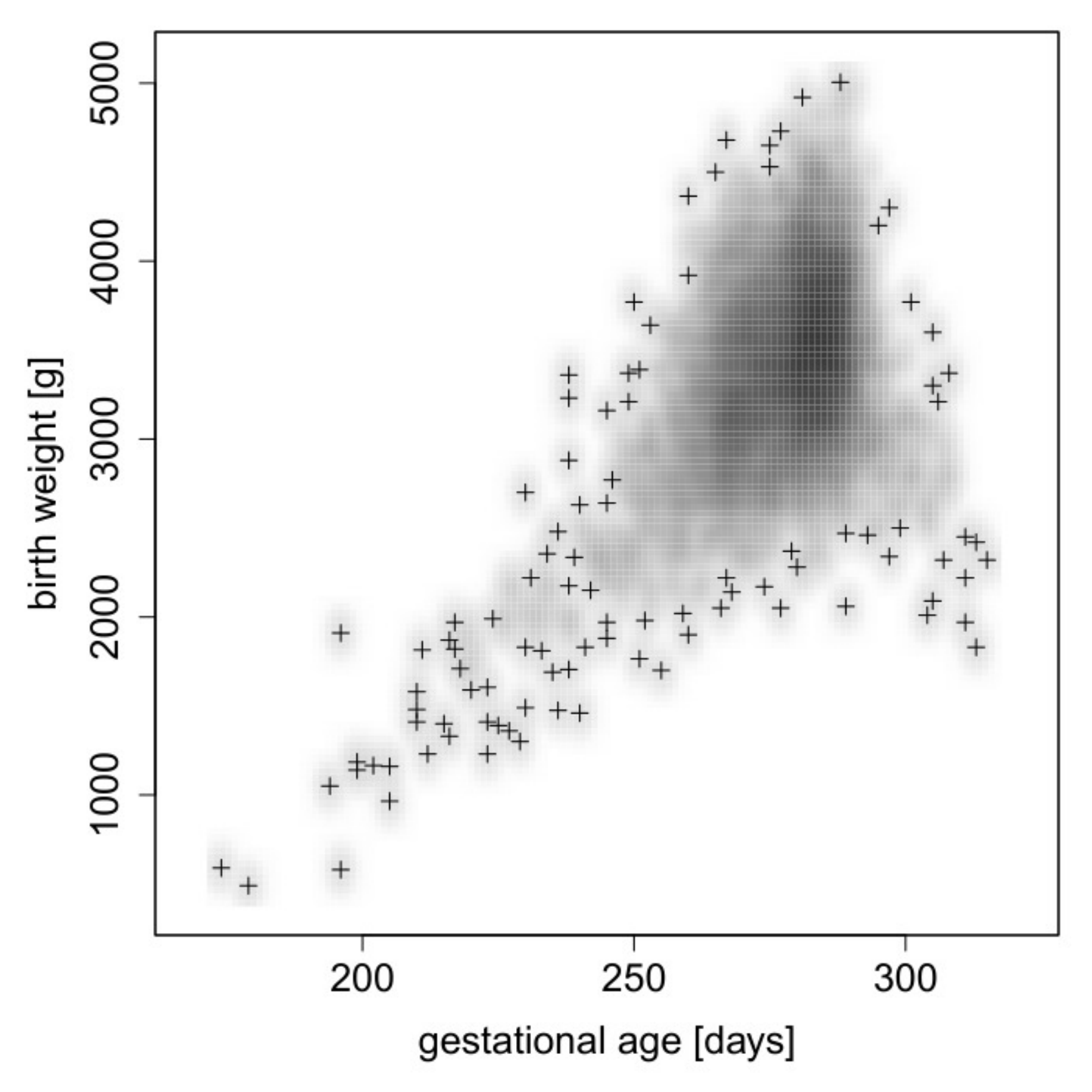}
\end{center}
\end{figure}

The response variables are BW (measured in \si{\gram} with varying accuracy, mean: 3390, standard deviation: 517) and GA (clinically estimated, in days, mean:~277, standard deviation:~12), the former being of primary interest (Figure~\ref{fig:resp}). Individual relevant covariates are pre-selected from the perinatal registry data. This is done in accordance with the literature (e.g., \citealp{Gar95, Sch10, Fre08, Tho01}), 
and with previous findings within our larger \textbf{PerSpat} project \citep{Kol16}. The specific variables are: child's sex,  number of previous pregnancies of the mother, whether the child has been delivered by Cesarean section (sectio), whether the birth has been induced, mother's age, mother's height, mother's body mass index (BMI) at the beginning of pregnancy, gain of weight of the mother during pregnancy, number of cigarettes the mother reports to smoke per day, whether the mother is single and whether the mother is employed. Some descriptive characteristics can be found in Table~\ref{tab:descr}. 
Temporal and spatial information is given by the date of birth and the postal code of the mother's residence. These are later used to include the average PFOA concentration in drinking water for a given place and time as a spatio-temporally assigned covariate (see Section~\ref{sec:expos}).

\begin{table}[htp]
\begin{center}
\caption{Descriptive characteristics for covariates from the perinatal data.}\label{tab:descr}
	\begin{tabular}{rlll}
	$j$ & Covariate & Unit &  Description  \\\hline
	1 & sex && female: 47\%  \\ 
	2 & previous pregnancies && 0: 38\%, 1: 32\%, 2: 16\% \\ 
	3 & sectio && 24\% \\ 
	4 & induction && 26\%  \\ 
	5 & maternal age &years& mean:~29.4, s.d.:~5.5 \\ 
	6 & maternal height &cm & mean:~167.0, s.d.:~6.7  \\ 
	7 & maternal BMI & kg/m$^2$ & mean:~25.2, s.d.:~5.3 \\ 
	8 & maternal gain of weight & kg & mean:~10.4, s.d.:~5.7  \\ 
	9 & maternal smoking & cigs./day & no: 87\%, $\le$10 cigs.: 8\%  \\ 
	10 & mother is single && 7\%   \\ 
	 $m=11$ & mother is employed &&  42\%  
	\end{tabular} 
\end{center}
\end{table}

\section{Bayesian Bivariate Distributional Copula Regression}\label{sec:cop}

To represent our data, let $n$ be the number of observations with bivariate responses $(y_{i1}, y_{i2})$, $i=1, \ldots, n$, from continuous response variables $Y_1$ and $Y_2$. We assume having $m$ covariates $x_{j}$, $j=1, \ldots, m$, with observations $x_{ij}$. We denote probability density functions $f_1$ and $f_2$ and cumulative distribution functions (CDFs) $F_1$ and $F_2$ of $Y_1$ and $Y_2$, respectively. 

In this section, we outline the employed statistical method of Bayesian conditional copula models within a distributional regression framework \citep{Kle16} with a focus on relevant components for our analysis. For a more general perspective, we refer to \cite{Kle16} and references therein.

\subsection{Copula and Marginal Distributions}

A bivariate copula is defined by a CDF $C_\rho : [0,1]\times[0,1] \to [0,1]$, such that the joint CDF of $Y_1$ and $Y_2$ can be written as $F(y_1, y_2) = C_\rho(F_1(y_1), F_2(y_2)) =: C_\rho(u, v).$ Sklar's theorem \citep{Skl59} ensures that $C_\rho$ always exists and is unique for continuous $Y_1$ and $Y_2$, while $F_1$ and $F_2$ are uniformly distributed. With copula density $c_\rho (\cdot, \cdot)$, the joint density of $Y_1$ and $Y_2$ can be written as $f(y_1, y_2) = c_\rho(F_1(y_1), F_2(y_2)) \cdot f_1(y_1) \cdot f_2(y_2)$ and a conditional density as 
\begin{equation}\label{eq:cond}
f_{1 | 2}(y_1 | y_2) = c_\rho(F_1(y_1), F_2(y_2)) \cdot f_1(y_1).
\end{equation}
While this representation is unconditional, the results can be extended to the regression context \citep{Pat06}.

There are various families of copulas, characterized by a  parameter $\rho$ representing the degree and form of dependence between $Y_1$ and $Y_2$. In our analysis, we compare the Gaussian copula family with density 
\begin{align*}
c_\rho (u, v) &= \dfrac{1}{\sqrt{1-\rho^2}}\exp\left( -\dfrac{1}{2}\cdot\dfrac{\rho}{1-\rho^2} \left(\rho(\Phi^{-1}(u))^{2}-2\cdot\Phi^{-1}(u)\cdot\Phi^{-1}(v)+\rho(\Phi^{-1}(v))^{2}\right) \right),\\
\rho &\in (-1,1),\\
\intertext{the Clayton copula family with density}
c_\rho (u, v) &= (1 + \rho)  (u  v)^{-1 - \rho}  \left(u^{-\rho} + v^{-\rho} - 1\right)^{-2 - 1 / \rho}\\
\rho &\in (0,+\infty),\\
\intertext{and the Gumbel copula family with density}
c_\rho (u, v) &= \dfrac{1}{uv} (-\ln u)^{\rho-1} (-\ln v)^{\rho-1} \exp\left(-z^{1 / \rho}\right) \left( z^{2 / \rho-2} - (1-\rho) z^{1 / \rho-2} \right),\\
& \text{where } z := (-\ln u)^{\rho} + (-\ln v)^{\rho},\\
\rho &\in (1,+\infty).
\end{align*}
As opposed to the Gaussian copula that allows for linear dependence and symmetry only, the Clayton copula allows for non-linear dependence between the two variables, in particular within the region of their extremely low values (tail dependence), whereas the Gumbel copula allows for upper tail dependence \citep{Kle16}. In the Gaussian case, the parameter $\rho$ represents the correlation between the two outcome variables. For the other copula models, a higher value of $\rho$ also signifies a stronger association between them. The copula parameter is also monotonically related to Spearman's Rho and Kendall's Tau \citep{Dal19}. For the latter, it holds explicitly: $\tau = \dfrac{\rho}{\rho + 2}$ for the Clayton and $\tau = 1 - \dfrac{1}{\rho}$ for the Gumbel copula \citep{Gha20}.

The two marginal distribution families can be chosen independently from each other and from the copula. Besides the Gaussian distribution $N(\mu, \sigma^2)$ with expected value $\mu$ and variance $\sigma^2$ for BW, a Dagum  distribution with density $f_{p,a,b}(y) = \dfrac{ap}{y}  \cdot\dfrac{\left(y/b\right)^{ap}}{\left( \left(y/b\right)^{a} + 1 \right)^{p+1}},$ shape parameters $p>0$ and $a>0$ and dispersion parameter $b>0$ is considered for GA in our study, see Section~\ref{sec:marg_choice} for more details on these choices.

\subsection{Resulting Joint Distributional Regression}\label{sec:regr}

All this is considered conditional on covariates by \cite{Kle16}, within the framework of structured additive regression models \citep{Fah04}. For any parameter $\theta$ of the joint distribution of $Y_1$ and $Y_2$,  being either one of the assumed marginal distributions or of the copula (i.e., $\theta \in \{\mu,\sigma^2, p,a,b, \rho\}$ in our case), a linear predictor $$\eta^{(\theta)} = \beta_0^{(\theta)} + \beta_1^{(\theta)} x_1 + \ldots + \beta_m^{(\theta)} x_m$$ is formed from the covariates numbered $j = 1, \ldots ,m$ (see Table~\ref{tab:descr}), possibly  just from a part of them or even reduced to the intercept. Link functions $h_\theta$ such that $\theta = h_\theta^{-1}(\eta^{(\theta)})$  
are specified appropriately for the respective parameter spaces: 
$\mu = \eta^{(\mu)}$, $\theta = \exp(\eta^{(\theta)})$ for $\theta \in \{\sigma^2, p,a,b\}$, $\rho = \dfrac{\eta^{(\rho)}}{\sqrt{1+(\eta^{(\rho)})^2}}$ for the Gaussian copula, $\rho = \exp(\eta^{(\rho)})$ for the Clayton copula and $\rho = \exp(\eta^{(\rho)})+1$ for the Gumbel copula.

The covariates to be included to the linear predictor $\eta^{(\theta)}$ can be separately selected for all parameters $\theta \in \{\mu,\sigma^2, p,a,b, \rho\}$. We consider those listed in Table~\ref{tab:descr}, without interactions. 

All models are estimated using a developer version of the \texttt{BayesX} software \citep{BayesX}, which implements fully Bayesian inference based on MCMC simulation techniques, see \cite{Kle16} for details.

\section{Bivariate Model Fitting to Perinatal Registry Data}\label{sec:choice}

We apply distributional copula regression models (Section~\ref{sec:cop}) to our perinatal registry data. We evaluate the models using the anonymized excerpt of $n=4451$ observations representing the five postal code areas of the town of Arnsberg in the upper Ruhr region, North Rhine-Westphalia, Germany, for the years of 2003 until 2014 (Section~\ref{sec:project}). Besides the \texttt{BayesX} software, calculations have been performed using the \texttt{R} environment \citep{R}, with the Dagum distribution from the \texttt{VGAM} package \citep{Yee20} and copula distributions from \texttt{copula} \citep{Hof20}.

\subsection{Preparation and Outline of Procedure}\label{sec:prep}

After preparation steps of data import and cleansing, the response data of observations $i=1, \ldots, n$ are standardized for numerical reasons only, without affecting the results, since the original responses can easily be recovered by linear back-transformation. Specifically, to employ a Gaussian for the marginals in \texttt{BayesX}, we use data-independent values for mean and standard deviation in a reasonable scale to yield $\tilde{y}_{i1} := \dfrac{y_{i1} - 3500}{500}$ for the birth weight (BW) and $\tilde{y}_{i2} := \dfrac{y_{i2} - 280}{14}$ for the gestational age (GA). To apply the Dagum marginal (with positive support), BW is normalized to $\tilde{y}_{i1} := \dfrac{y_{i1}}{500}$, while GA is also inverted to a more appropriate shape by $\tilde{y}_{i2} := \dfrac{322 - y_{i2}}{14}$, to have the main part of the data closer to zero and the tail on the right (322 days $=$ 46 weeks exceed the maximum observable GA). 

\begin{figure}[hp]
\caption{Outline of procedure to choose optimal marginal and copula families in bivariate Bayesian distributional regression.}\label{fig:flow}
\begin{center}
	\includegraphics[width=.98\textwidth]{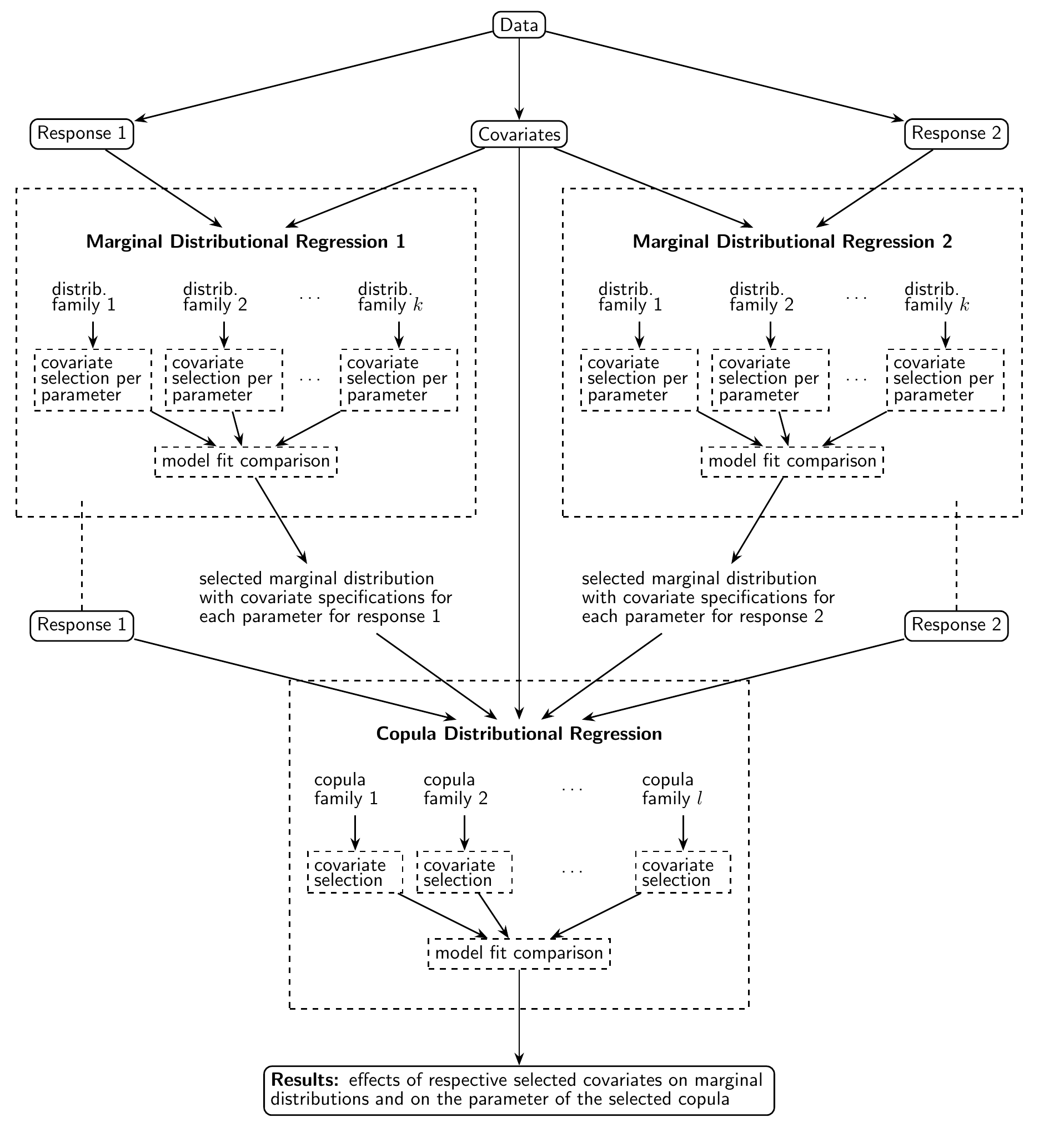}
\end{center}
\end{figure}

We choose the optimal copula regression model in a stepwise procedure, outlined in Figure~\ref{fig:flow}. The copula property (Section~\ref{sec:cop}) enables  separate considerations of the marginal distributions of BW and GA and of their dependence structure. This motivates to identify optimal marginal model fits first and to apply them in the search for the best fitting copula model afterwards. 
Variable selections within this model choice process help to ease later evaluation steps and give a first insight in the relevance of covariates; however, the additional uncertainty has to be kept in mind when considering final results in terms of significance. 

Marginal distribution families are chosen by applying Gaussian and Dagum models to both univariate responses; in all these four models and for all parameters therein ($\theta \in \{\mu, \sigma^2\}$ or $\theta \in \{p, a, b\}$, respectively), covariates are selected if the 95\%-credible-intervals of their respective $\beta_j^{(\theta)}$'s do not include zero; the optimal models per family are compared by probability integral transform values, quantile residuals and log-scores (Section~\ref{sec:marg_choice}). The optimal marginals of BW and GA are combined with all possible copula families (including rotations); covariates for the copula parameter are selected and these models compared by information criteria (Section~\ref{sec:cop_choice}). The final model is evaluated in terms of prediction performance and substantive results (Section~\ref{sec:res}).

\subsection{Details on Marginal Specifications}\label{sec:marg_choice}

After excluding non-significant covariates, i.e., where $0 \in \text{CI}_{95\%}\left(\beta_j^{(\theta)}\right)$, different univariate distributional models \citep{Kle15} are fitted and compared using logarithmic scores (log-scores,  \citealp{Gne07}). To compute the log-scores, we implement a four-fold cross-validation, for which the observations are randomly assigned to subsamples of equal size. 
Using the estimated model based on three subsamples, individual log-scores for the respective left-out subsample are computed using the \texttt{R} package \texttt{scoringRules} \citep{Jor19}. The average log-scores for Dagum and Gaussian distribution are very close to each other in the case of BW (Dagum: 7.63, Gaussian: 7.66); however, for GA, the Dagum distribution has a notably better fit (3.74 vs. 4.07).

\begin{figure}[ht]
\caption{Histograms of probability integral transform values $F(y; \hat{\beta})$ (with posterior means of the $\beta$'s plugged in) for the two models and the two marginal response variables.}\label{fig:pit}
\begin{center}
	\includegraphics[width=.49\textwidth]{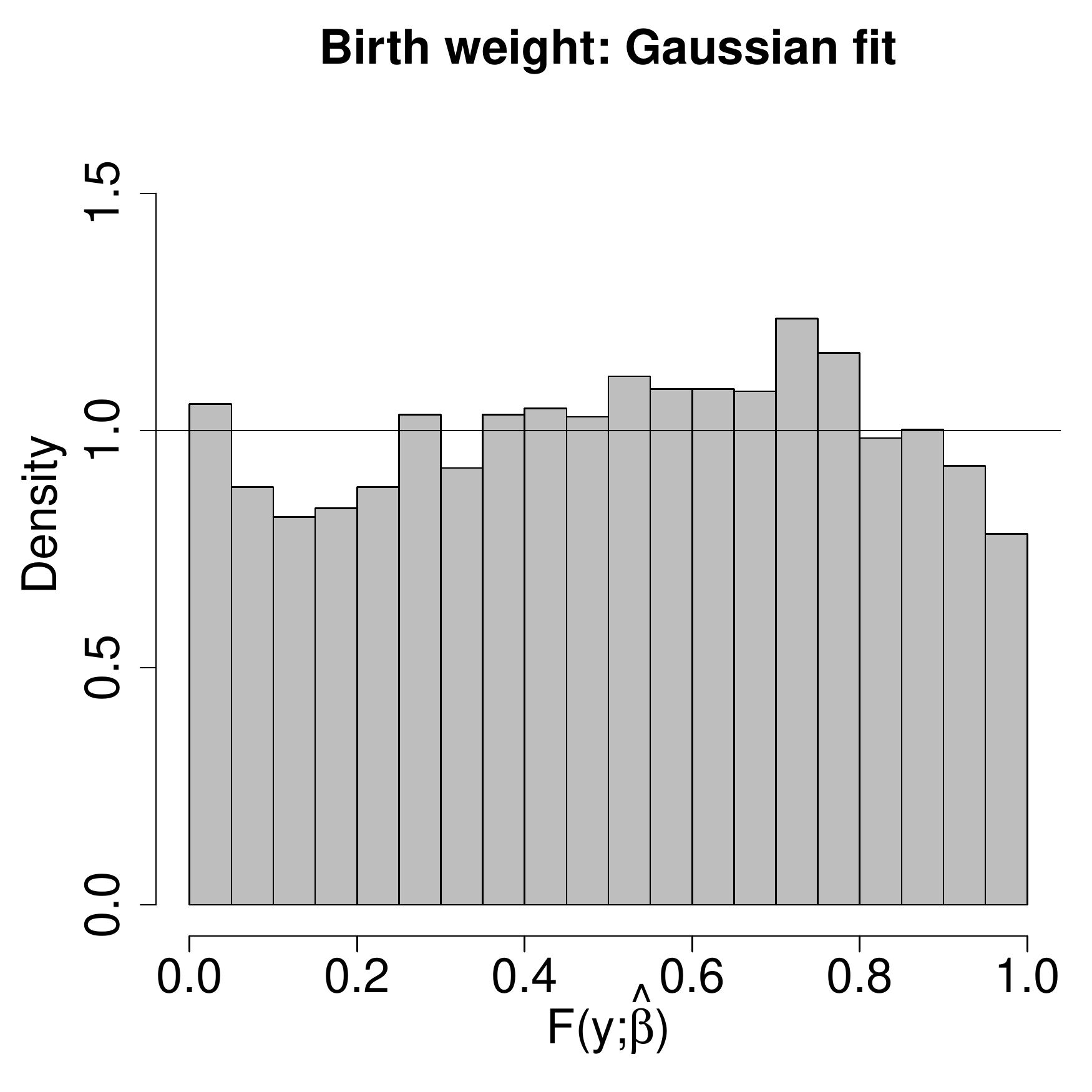} 
	\includegraphics[width=.49\textwidth]{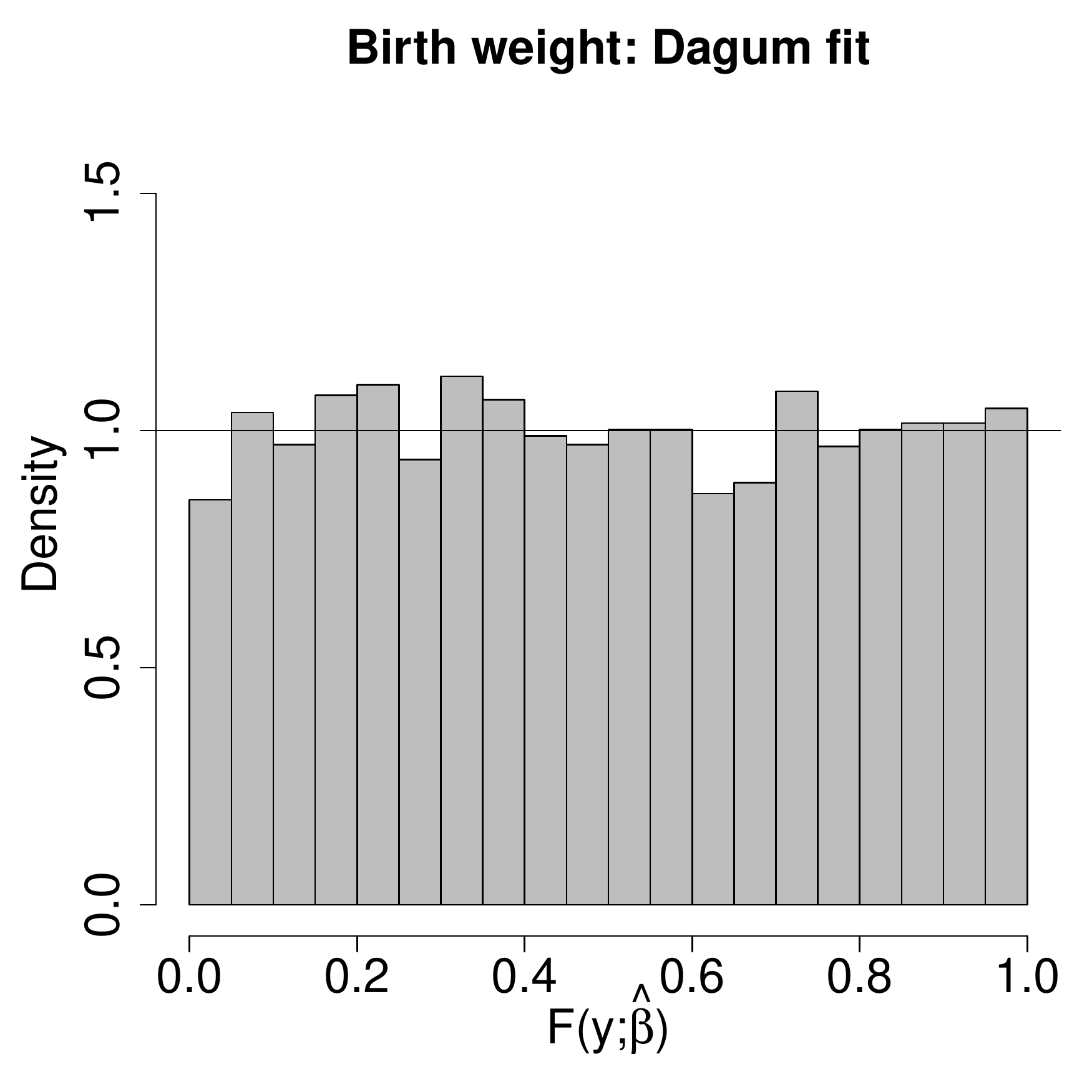} \\
	\includegraphics[width=.49\textwidth]{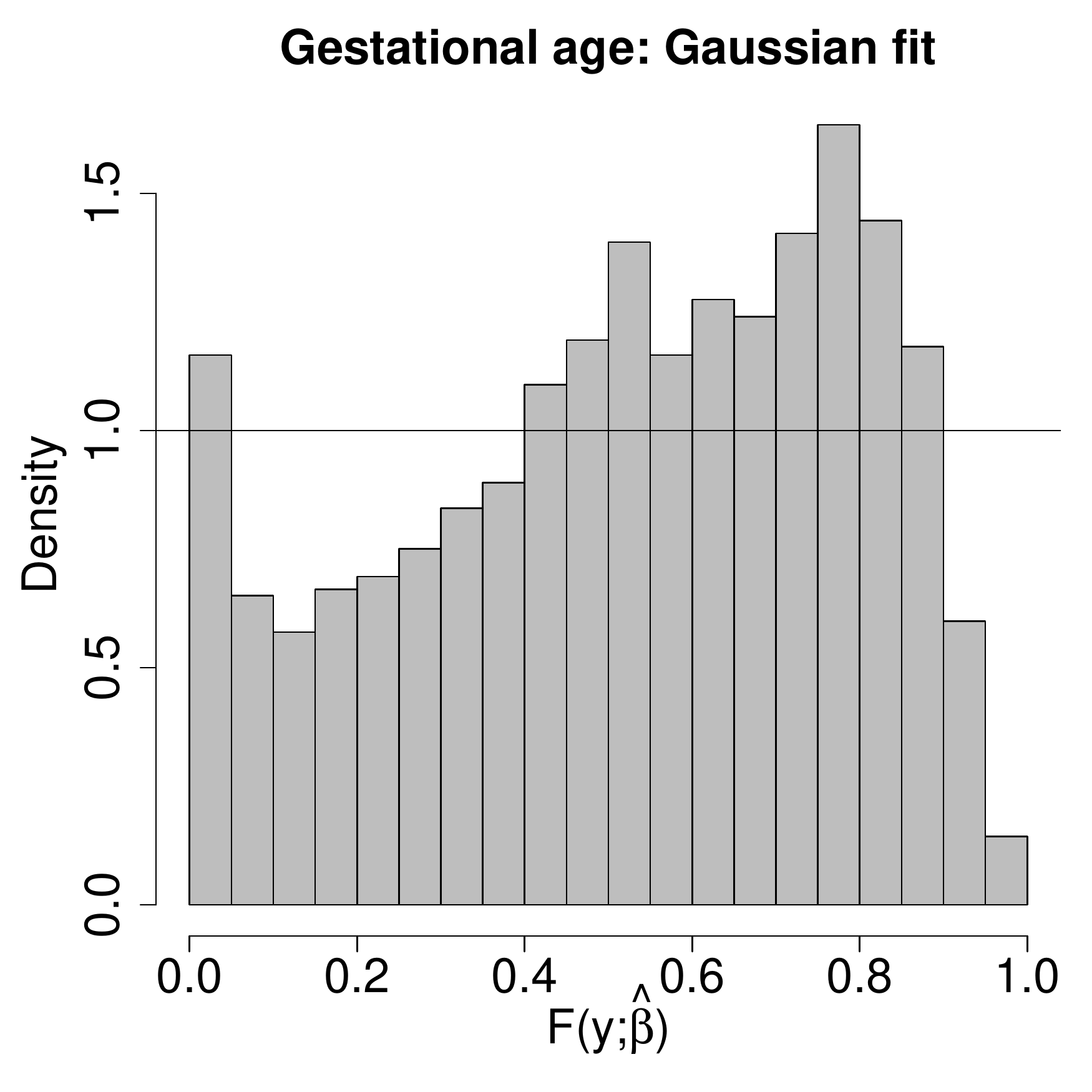} 
	\includegraphics[width=.49\textwidth]{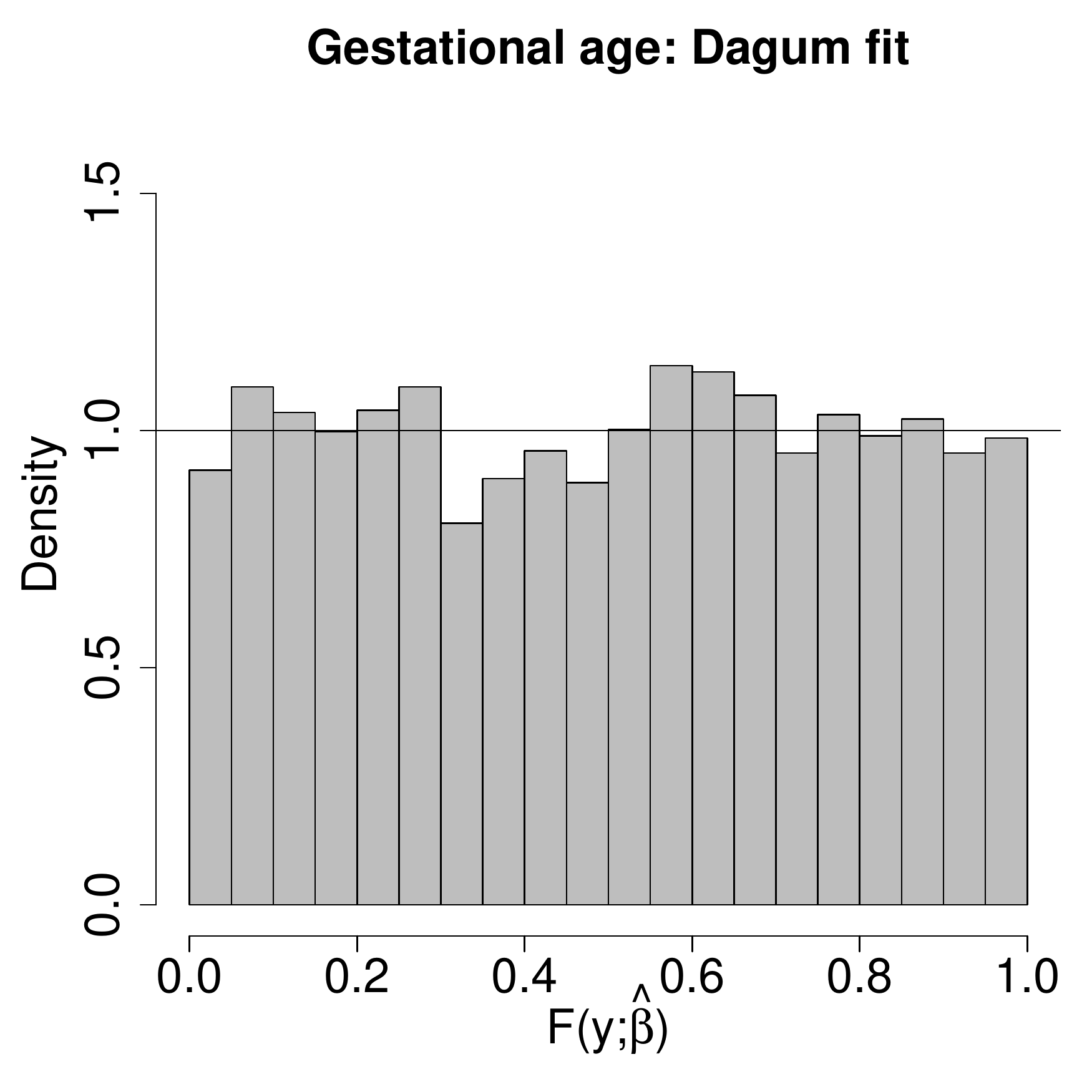}
\end{center}
\end{figure}

\begin{figure}[ht]
\caption{Quantile-quantile-plots of randomized quantile residuals $\Phi^{-1}(F(y; \hat{\beta}))$ (with posterior means of the $\beta$'s plugged in) for the two models and the two marginal response variables.}\label{fig:qq}
\begin{center}
	\includegraphics[width=.49\textwidth]{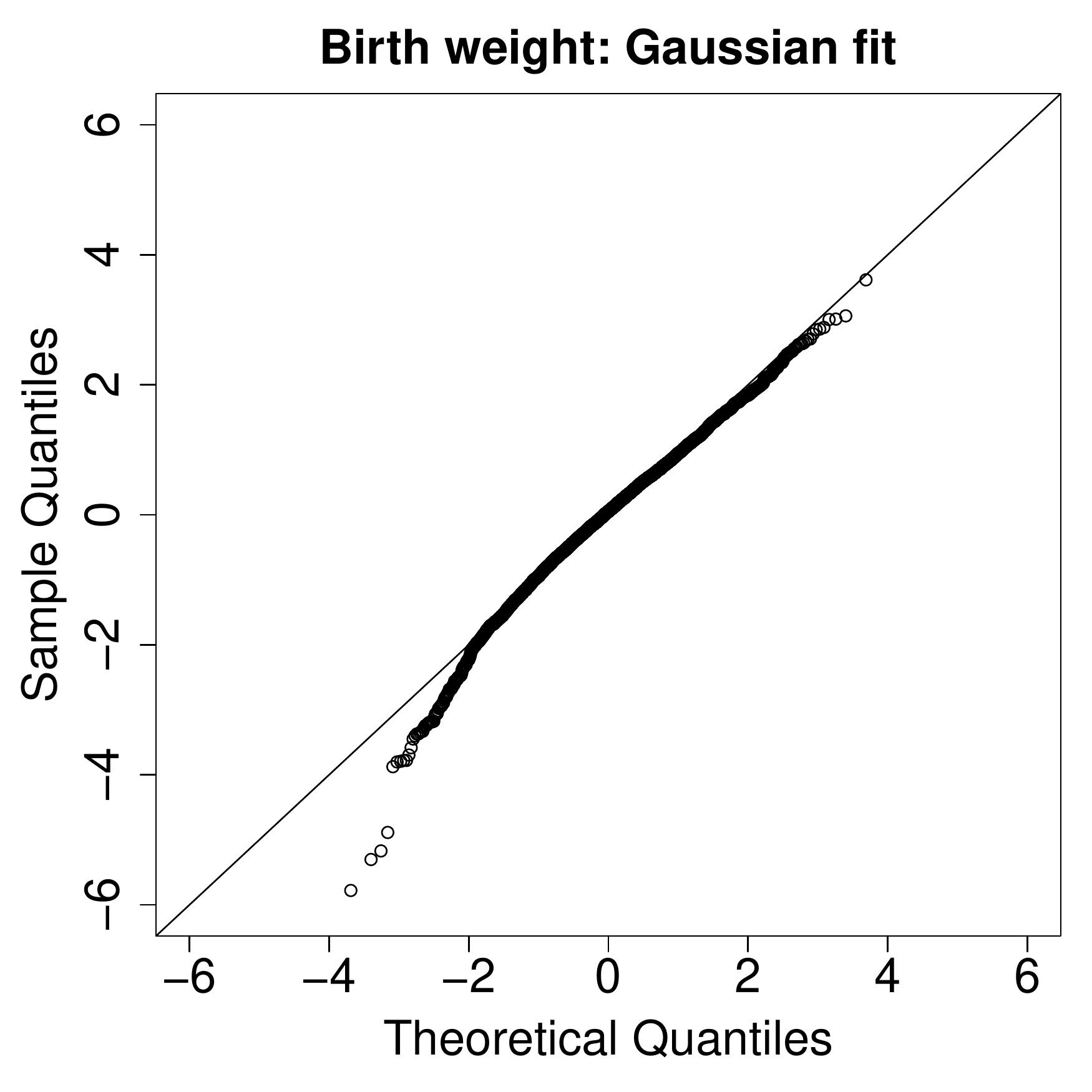} 
	\includegraphics[width=.49\textwidth]{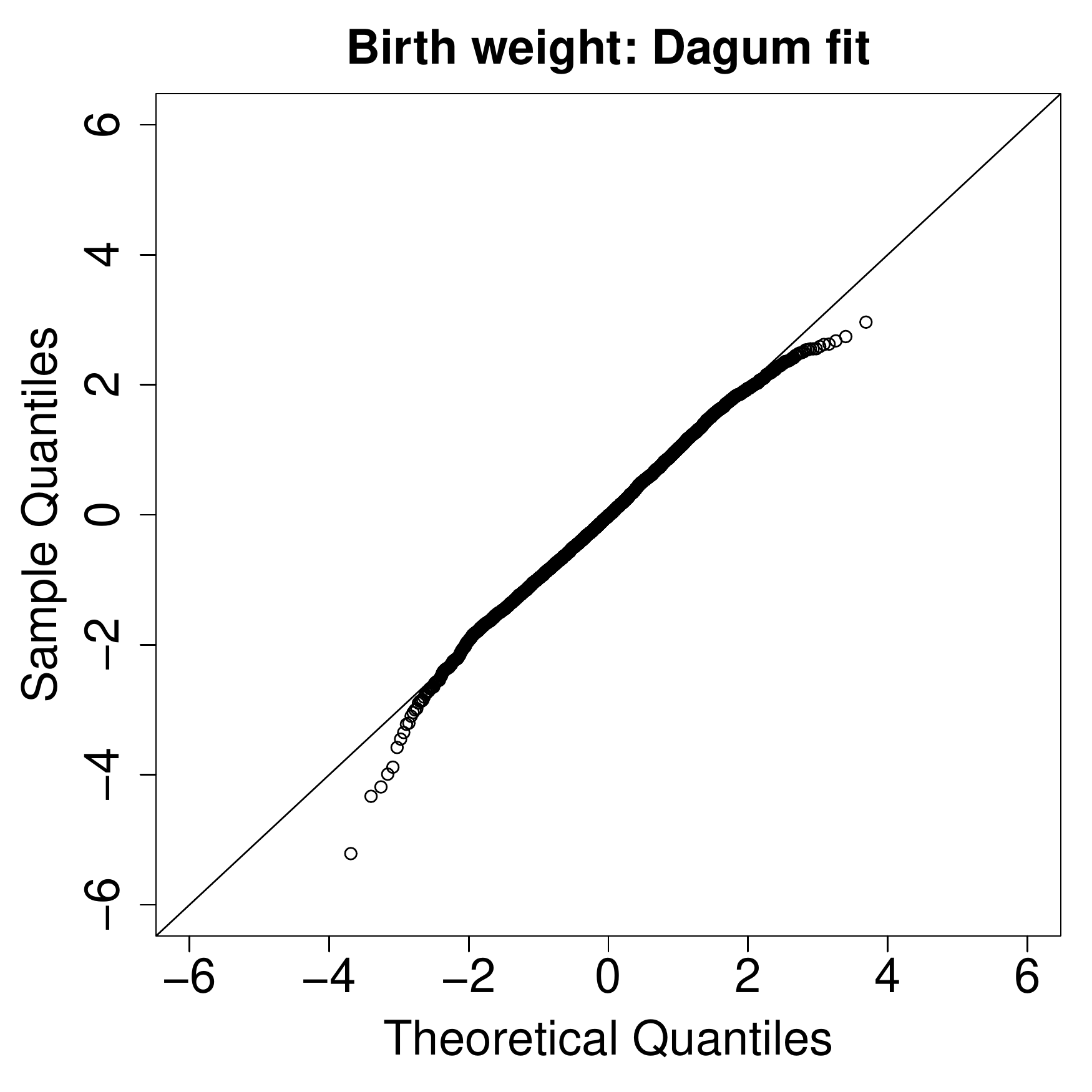} \\
	\includegraphics[width=.49\textwidth]{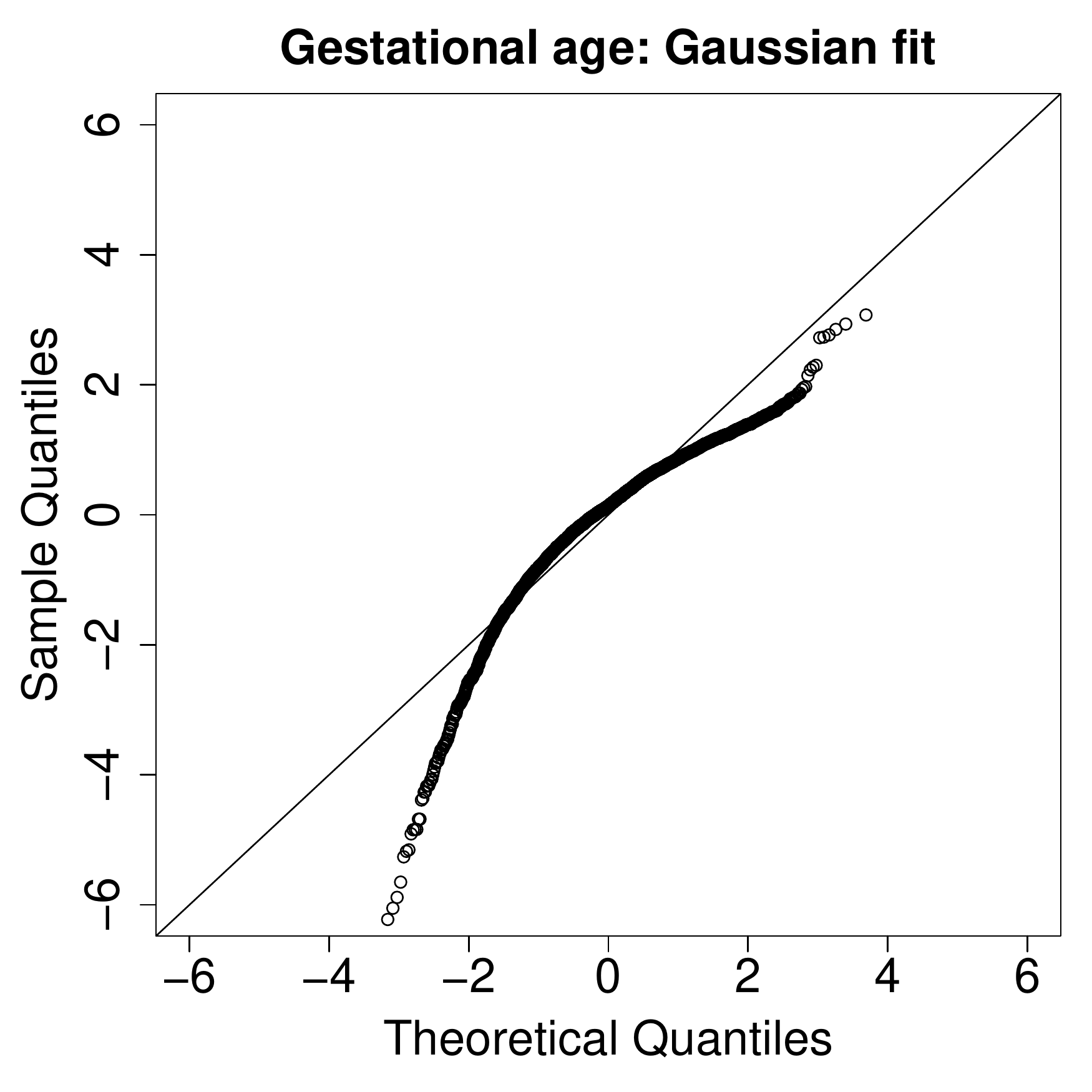} 
	\includegraphics[width=.49\textwidth]{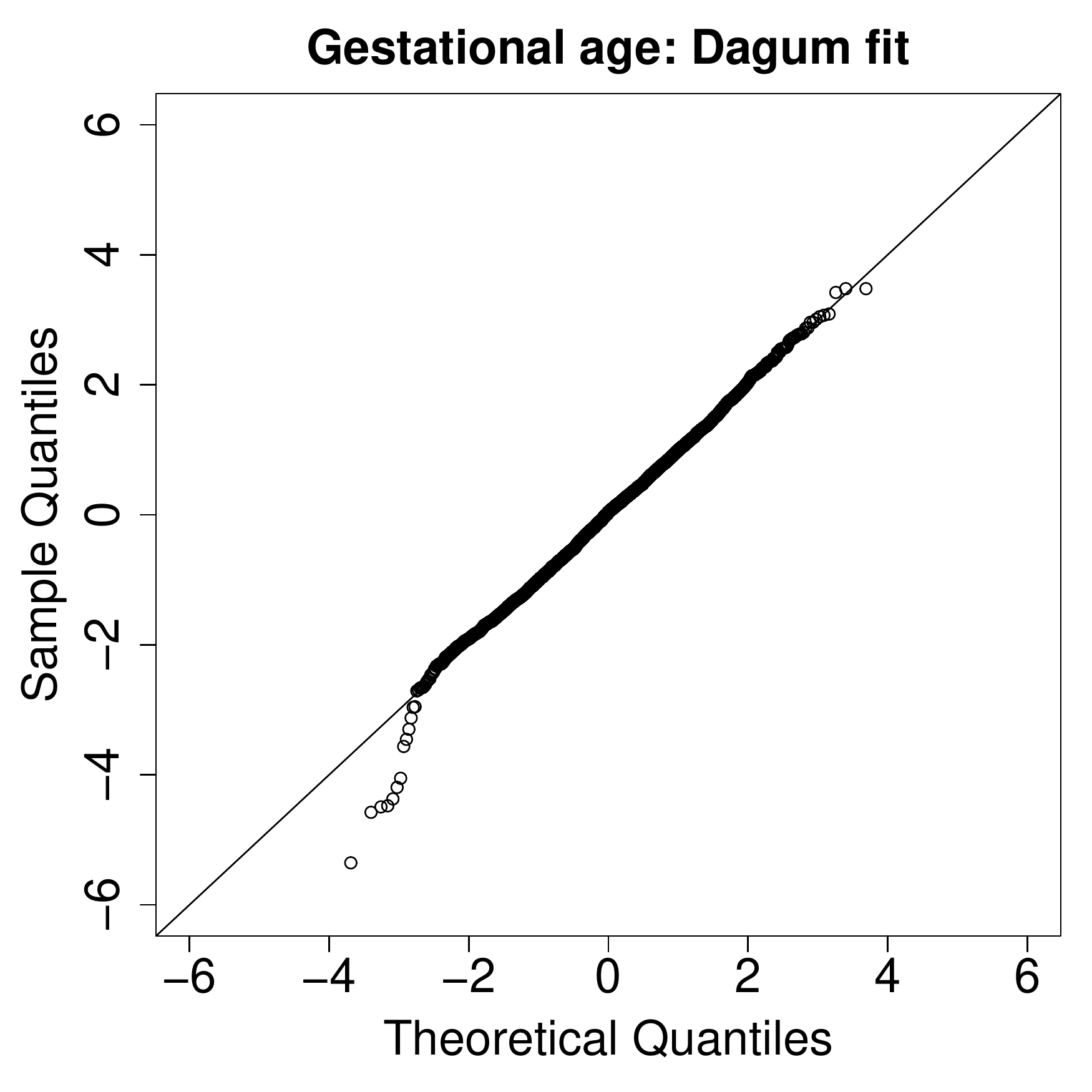}
\end{center}
\end{figure}

These findings are confirmed by graphical evaluation of the probability integral transform values $F(y; \hat{\mathbf{\beta}})$ and the corresponding normalized quantile residuals $\Phi^{-1}(F(y; \hat{\mathbf{\beta}}))$ (cf., \citealp{Dun96}), where posterior means of the respective $\beta_j^{(\theta)}$'s are employed and $y$ (BW or GA) is on the standardized scale. The theoretically expected uniform distribution of the $F(y; \hat{\mathbf{\beta}})$ is well recognisable for the Dagum fits (Figure~\ref{fig:pit}), 
while it is strongly violated for the Gaussian fit of GA; a similar structure as for the latter remains also for the Gaussian fit of BW, but considerably weaker.  A quantile-quantile-plot of the $\Phi^{-1}(F(y; \hat{\mathbf{\beta}}))$ (Figure~\ref{fig:qq}) shows a slightly better fit of the Dagum model in either case, especially in the lower range, but more striking for GA, where a distinguishable structure remains for the Gaussian. These results are convincing to use the Dagum distribution for the GA marginal in the further analyses.

For BW, however, the results are less clear and there are reasons to stick with the Gaussian, in doubt: In this application, we are primarily interested in influences of covariates on mean (and variability) of BW -- characteristics directly represented by the $N(\mu, \sigma^2)$ parametrization; and there is at least some evidence above that this distribution family does not fit essentially worse than the other. In this parametrization,  effects of covariates are easily and directly interpretable. By contrast, the interpretation of effects on the three Dagum parameters is quite complicated and sometimes ambiguous in terms of substantive results (cf., Section~\ref{sec:eval}). Results from the Gaussian fit are furthermore directly comparable to other studies, especially with standard regression models (cf., Section~\ref{sec:comp}). So, we use the Gaussian for the BW marginal in the further analyses.

\subsection{Details on Copula Specification}\label{sec:cop_choice}

\begin{figure}[ht]
\caption{Rank correlations of birth weight and gestational age (80\%- and 95\%-confidence-intervals) from subsets of data selected by certain covariate choices: left: Caesarian section, two groups; right: maternal gain of weight, ordered observations grouped into four equal subsets; other covariates do not feature such visible differences between subsets.}\label{fig:condcor}
\begin{center}
	\includegraphics[width=.49\textwidth]{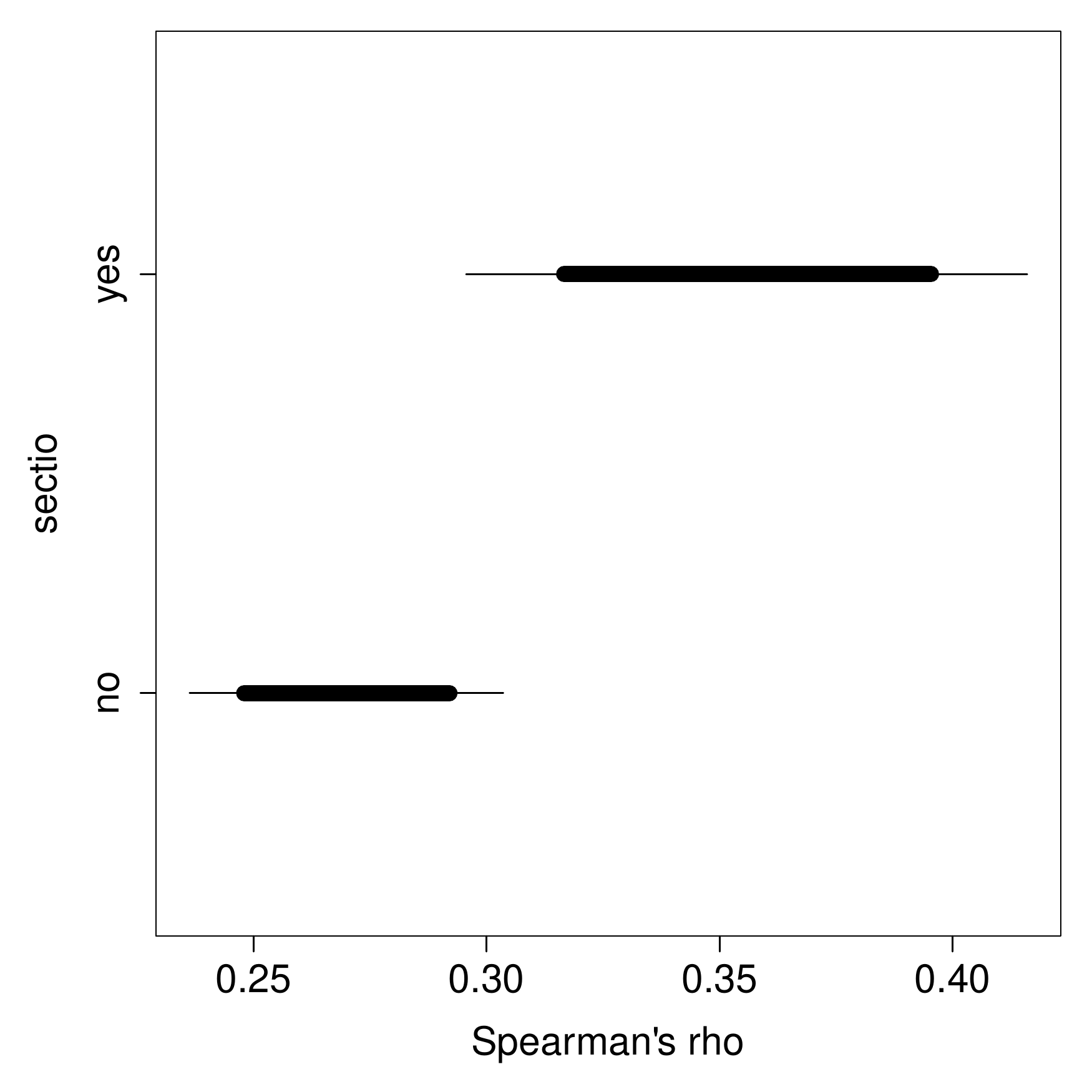}
	\includegraphics[width=.49\textwidth]{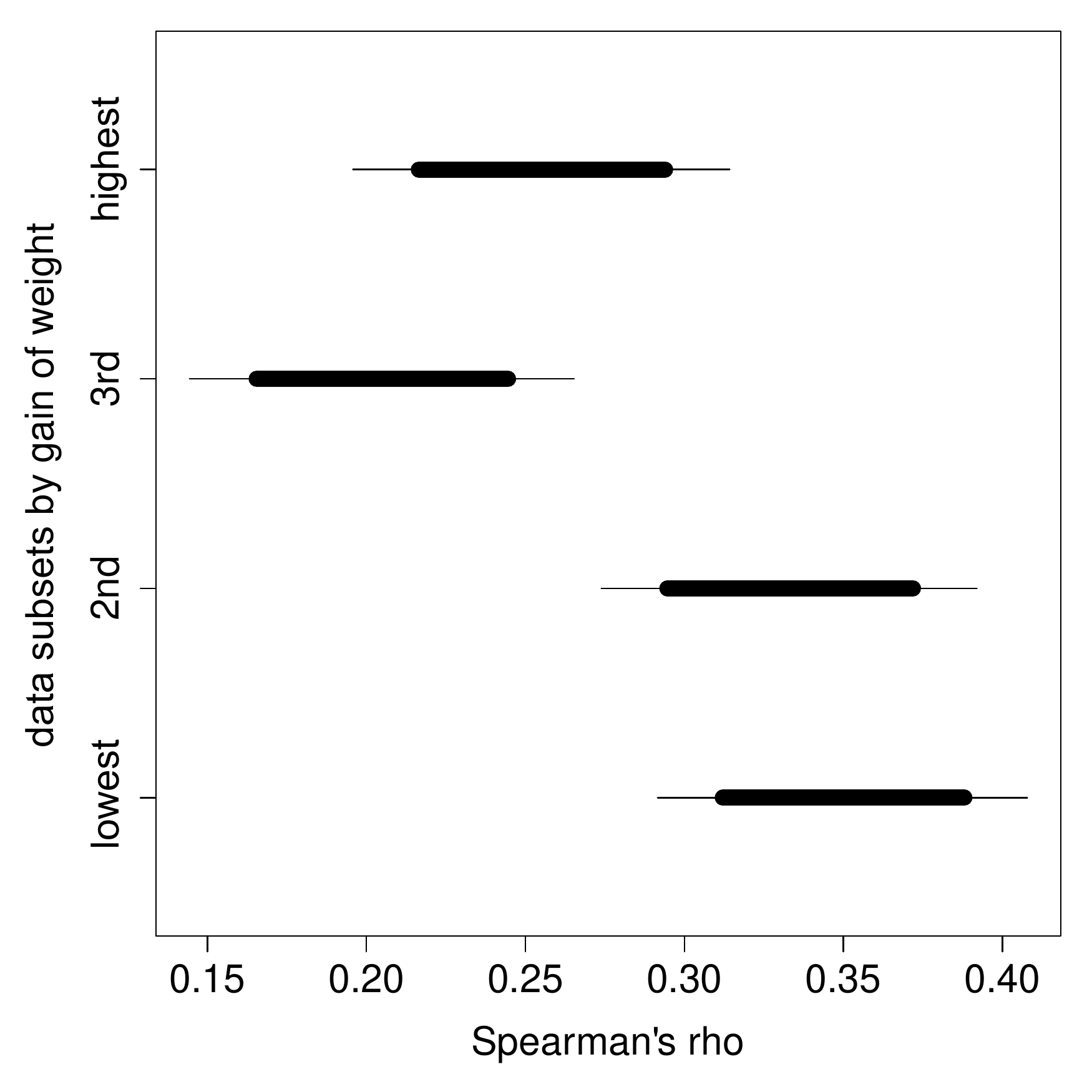}
	\end{center}
\end{figure}

As a motivation, we illustrate the usefulness of flexible copula regression and non-linear dependence structures by comparing confidence intervals of rank correlation coefficients (\texttt{R} package \texttt{spearmanCI}, \citealp{Car18}) of BW and GA from subsets of data selected by certain covariate choices: In case of a discrete covariate, we consider the respective levels; in case of a continuous covariate, observations are ordered by the covariate values and grouped together into four subsets of equal size. Differences between these levels or subsets are especially pronounced for the sectio and weight-gain covariates (Figure~\ref{fig:condcor}). 

Therefore, we compare the Gaussian, Clayton, and Gumbel copula using the deviance information criterion (DIC, \citealp{Spi02}) and the widely application information criterion (WAIC, \citealp{Wat10}). Since too many covariates with respect to the $\rho$ parameter of the Clayton and Gumbel copulas lead to 
software crashes with our amount of data, we pre-select them based on the variability of correlation coefficients reported above and by tentatively adding covariates one by one. With respect to the three copula families, the Clayton copula model yields the best DIC and WAIC values, which is also in line with the descriptive picture from Figure~\ref{fig:resp}.

In conclusion from this model fitting section, we use a Clayton copula with Dagum marginal for GA and Gaussian marginal for BW. The predictor specifications for each of the six model parameters $\rho$, $p$, $a$, $b$, $\mu$ and $\sigma^2$ are given in Table~\ref{tab:linpred}, the respective link functions specified in Section~\ref{sec:regr} are employed. In MCMC sampling, three chains are run to double-check the results. The respective samples from the posterior appear to be from the same distribution, leaving but small potential scale reduction factors (obtained using the \texttt{R} environment following \citealp{Bro98}, Table~\ref{tab:psrf}). The effective sample sizes (obtained using the \texttt{R} package \texttt{CODA}, \citealp{CODA}, Table~\ref{tab:ess}) signify that there is mostly no autocorrelation detectable in the posterior samples. These two measures per regression parameter are available in the Appendix.

\begin{table}[ht]
\begin{center}
\caption{Final bivariate copula model specification and result overview: composition of the linear predictor $\eta^{(\theta)}$ from the covariates, per parameter $\theta \in \{\mu,\sigma^2, p,a,b, \rho\}$ of the chosen marginal and copula families, together with the employed link functions. Included covariates are marked by $+$, $-$ or $\circ$, where $+$ and $-$ denote significant positive and negative effects, respectively, and $\circ$ is for no significant effect in the final evaluation.}\label{tab:linpred}
\begin{tabular}{cl|cc|ccc|c}\\
& \textbf{Response} &  \multicolumn{2}{c|}{Birth Weight} &  \multicolumn{3}{c|}{Gestational Age}  & \textbf{Copula} \\ 
& \textbf{Family} & \multicolumn{2}{c|}{Gaussian} &  \multicolumn{3}{c|}{Dagum} & Clayton \\ 
& \textbf{Parameter} & $\mu$ & $\sigma^2$ & $p$ & $a$ & $b$ & $\rho$ \\\hline
 \multirow{11}{*}{\begin{sideways}\textbf{Components of }$\eta^{(\theta)}$\end{sideways}} & sex (female) & $-$ & $-$ &   &   &   &  \\ 
  & prev. pregn. & $+$ &  &   &   & $+$ &  \\ 
  & sectio & $-$ & $+$ & $-$ & $-$ & $+$ & $+$ \\ 
  & induction & $+$ &   & $+$ & $-$ & $-$ &  \\ 
  & mat. age &  &   &   &   &   &  \\ 
  & mat. height & $+$ &   &   &   &   &  \\ 
  & mat. BMI & $+$ & $+$ &   &   &   &  \\ 
  & weight gain & $+$ &   &   & $+$ &   &  \\ 
  & smoking & $-$ & $+$ &   & $-$ &   &  \\ 
  & single & $\circ$ &   &   &   &   &  \\ 
  & employed &  &   & $\circ$ &   &   &  \\\hline
 \multicolumn{2}{r|}{\textbf{Link:} $\eta^{(\theta)} = \ldots $} & $\mu$ & $\ln\sigma^2$ & $\ln p$ & $\ln a$ & $\ln b$ & $\ln \rho$ \\ 
 \end{tabular}  
\end{center}
\end{table}

\section{Analysis of Perinatal Registry Data}\label{sec:res}

Based on the results from Section~\ref{sec:choice}, we apply the bivariate distributional copula regression model  to the perinatal registry data (Section~\ref{sec:eval}), following the procedure outlined in Figure~\ref{fig:flow}, which leads to the variable selection specified in Table~\ref{tab:linpred}. A standard univariate regression approach for birth weight (BW) is set up and performances of both models are compared (Section~\ref{sec:comp}). Finally, the local Perfluorooctanoic Acid concentrations in drinking water are integrated to the models (Section~\ref{sec:expos}). Our study is the first to apply a distributional copula regression model in this scenario.

\subsection{Evaluation of Copula Regression Model}\label{sec:eval}

Significant influences of covariates on BW's mean are quantified in Table~\ref{tab:coeff}. Apart from this, the distributional copula regression model reveals covariate-dependence of all the other parameters of the two-dimensional outcome, significant influences are summarized in Table~\ref{tab:linpred}. BW's scale ($\sigma^2$) is higher for male children, in case of sectio, for higher maternal BMI and if the mother smokes. 

\begin{table}[htp]
\begin{center}
\caption{Regression coefficients (posterior mean and standard deviation) regarding the parameter $\mu$ of (standardised) BW, estimated in the polynomial and the distributional copula regression model.}\label{tab:coeff}
	\begin{tabular}{lrrrr}
{Covariate} & \multicolumn{4}{c}{{Coefficients in regression models}}\\
	 & \multicolumn{2}{c}{Polynomial} & \multicolumn{2}{c}{Copula}  \\\hline
 sex (female) & $-$0.2908 & ($\pm$ 0.0237) & $-$0.2896    & ($\pm$ 0.0273) \\ 
previous pregnancies & 0.0583    & ($\pm$ 0.0085) & 0.0484    & ($\pm$ 0.0065) \\ 
 sectio & \multicolumn{2}{c}{not signif.} & $-$0.2907    & ($\pm$ 0.0535) \\ 
induction & \multicolumn{2}{c}{not signif.} & 0.0870 & ($\pm$ 0.0280) \\ 
maternal age & \multicolumn{2}{c}{not signif.} & \multicolumn{2}{c}{not signif.} \\ 
 maternal height & 0.0279    & ($\pm$ 0.0018) & 0.0290 & ($\pm$ 0.0015) \\ 
 maternal BMI & 0.0302    & ($\pm$ 0.0023) & 0.0368 & ($\pm$ 0.0029) \\ 
 maternal gain of weight  & 0.0144    & ($\pm$ 0.0022) & 0.0249    & ($\pm$ 0.0024) \\ 
 maternal smoking & $-$0.0308    & ($\pm$ 0.0029) & $-$0.0416    & ($\pm$ 0.0031) \\ 
 mother is single & $-$0.1271    & ($\pm$ 0.0480) &
	 \multicolumn{2}{c}{not signif.}  \\ 
 mother is employed & \multicolumn{2}{c}{not signif.} &
	 \multicolumn{2}{c}{not signif.}  \\\hline 
 gestational age & $-$2.6145   & ($\pm$  0.1717) &  &  \\ 
 squared gestational age & 0.0112    & ($\pm$ 0.0007) &  &  \\ 
	cubic gestational age & $-$15\textsc{e}$-$6    & ($\pm$ 0.9\textsc{e}$-$6) &  &  
	\end{tabular} 
\end{center}
\end{table}

For the Dagum distribution of gestational age (GA), the shape parameter $p$ is higher in case of induction and lower in case of sectio. The shape parameter $a$ increases with the maternal gain of weight and is lower in case of sectio, induction and if the mother smokes. The scale parameter $b$ increases with the number of previous pregnancies and in case of sectio, and is lower in case of induction. If we consider the distribution's median $b\cdot\left(-1 + 2^{1/p}\right)^{-1/a}$, the monotonically increasing link functions and the inverting transformation $\tilde{y}_{i2} = \dfrac{322 - y_{i2}}{14}$ of the data, we can qualitatively interpret these results such that GA is higher for decreasing $p$ or $b$ or increasing $a$, e.g., with increasing maternal gain of weight. But we also see that this interpretation is generally rather difficult. It leads to no consistent results in the case of sectio or induction. 

For the copula parameter representing the degree of dependence between BW and GA, only the information, whether the child has been delivered by Cesarean section, emerges as a stably estimated significant influence. Taking the intercept into account, the dependence between BW and GA measured in this way turns out to be surprisingly weak, in fact not far from independence ($\rho\approx 0.40$, 95\%-CI: $[0.21, 0.76]$, for children delivered by Cesarean section; $\rho\approx 0.14$, 95\%-CI: $[0.09, 0.22]$, for the others), although significantly positive ($\rho\searrow 0$ would signify independence).

\subsection{Univariate Polynomial Regression and Comparison to the Copula Approach}\label{sec:comp}

Instead of bivariate regression for BW and GA, univariate analyses for one of them are common in gynecological and obstetric research (e.g., \citealp{Skj00, Wei14, Fre08}), perhaps adjusted for the other, or with a dichotomous response like `small for GA' (e.g., \citealp{Pol09, Tho01}).

\subsubsection*{Standard Model Development}

In preliminary studies, we confirm a regression model as the most suitable among univariate BW models, where GA is included as a covariate in the form of a polynomial $p_\gamma$ of degree three: We apply polynomial regression models $y_{i1} = \beta_0 +  p_\gamma(y_{i2}) + \sum_j \beta_j x_{ij} + \epsilon_i$, $i = 1, \ldots, n$ with independent $\epsilon_i \sim N(0, \sigma^2)$ for BW, with observed GA $y_{i2}$ as covariate and some of the further covariates numbered $j = 1, \ldots ,m$ (see Table~\ref{tab:descr}). Among the usual fractional polynomials \citep{Roy08} of degree one or two, the resulting mean prediction errors are very close to each other. With regard to RSS, AIC, BIC and maximum prediction error (i.e., for outlying data), the polynomial $p_\gamma(y_{i2}) = \gamma_1 y_{i2}^2 + \gamma_2 y_{i2}^3$ performs best. However, a model with `full' polynomial $p_\gamma(y_{i2}) = \gamma_1 y_{i2} + \gamma_2 y_{i2}^2 + \gamma_3 y_{i2}^3$  is even better in this respect, is in accordance with gynecological and obstetric literature (e.g., \citealp{Gar95, Sal07}) and therefore preferred.

\subsubsection*{Model Fit Comparison}

We perform a four-fold cross-validation and compare the obtained prediction samples of BW from copula and standard model with the observed values by log-scores as in Section~\ref{sec:marg_choice}. 

For the copula model, after estimation of the bivariate joint distribution, we are interested in predictions of BW conditional on GA. To evaluate the conditional distribution with density~(\ref{eq:cond}), we draw random numbers via rejection sampling with a uniform envelope extended to a large enough range. Thereby, we use the observed GA values $y_{i2}$ and parameter estimates $\hat{\theta}$ obtained from samples of the posterior $\hat{\beta}_j^{(\theta)}$'s and the covariate values of the respective observations.

For the standard model, there results an average log-score of 7.41; for the copula model, with respect to BW response conditional on GA, it is 7.67. Thus, the standard model performs slightly better than the copula model. But it has to be noted, that the standard polynomial model is designed for this specific univariate situation and the dependence of BW and GA is expressed by three parameters $\gamma_1, \gamma_2, \gamma_3$. By contrast, the copula model is more general in the sense that it is intended for the bivariate response and its result is now reduced to the prediction of BW; it also includes only one parameter for the relationship of BW and GA. 

\begin{figure}[tp]
\caption{Top: residual plots for birth weight from polynomial (left) and copula distributional (right) regression model, each with smoothed mean and standard deviation lines; bottom right: the same for gestational age; bottom left: predictions of birth weight from polynomial and copula distributional regression model, plotted against each other per observation, with bisecting line (dotted) and robustly estimated principal axis of the plotted data (`direction of main point cloud', solid); predictions for all figures obtained from cross-validation study.}\label{fig:resid}
\begin{center}
	\includegraphics[width=.98\textwidth]{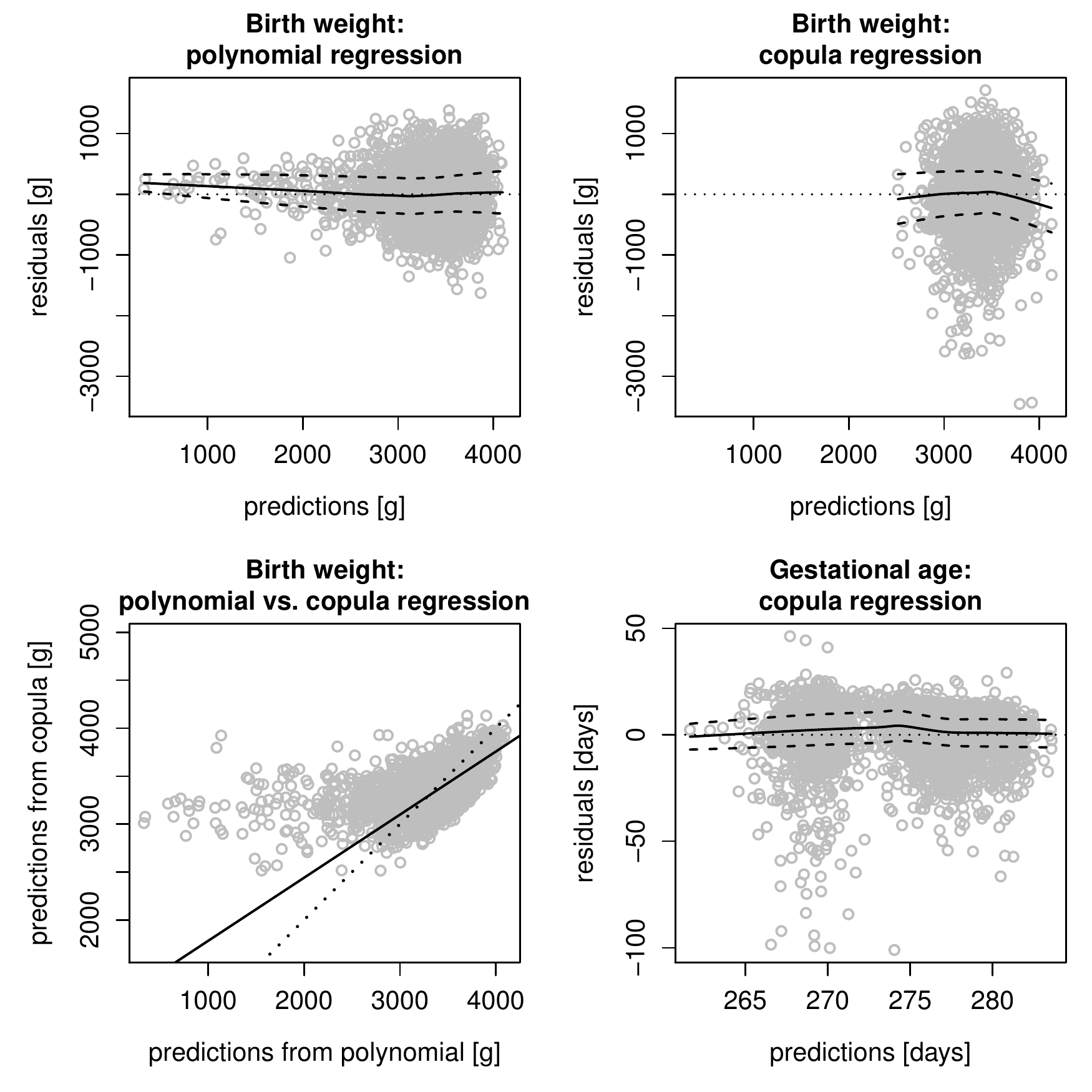}
\end{center}
\end{figure}

For the vast majority of BW predictions, the distributional copula regression model is close to univariate polynomial regression. 
The residual and comparison plots in 
Figure~\ref{fig:resid} show how the models agree, especially in mean (bottom left). However, extremely low BWs are correctly predicted by the polynomial alone (top left), while their observations diverge from the copula model predictions (top right, fitted values are in the range of the main part of data, but residuals are too far to the negative). A closer range of predictions from the copula model is also visible (top right). The residual plot for GA from copula regression (Figure~\ref{fig:resid}, bottom right) also reveals rather poor prediction of extremely low values, which often coincide with very low BW; besides, there emerge two distinguishable groups of  GA predictions, presumably in connection with the highly influential sectio and induction covariates.

Due to independent and simultaneous estimation of marginals and copula, estimates of the regression coefficients with regard to BW's mean are very similar in both models, but their significances differ (Table~\ref{tab:coeff}).

Substantive comparison of both models' results (Section~\ref{sec:disc}) further demonstrates the usefulness of bivariate distributional regression, in particular when the relationships of BW, GA and covariates are simultaneously analyzed to, e.g., reveal mediating effects.

\subsection{Integration of Local PFOA Drinking Water Concentrations}\label{sec:expos}

Perfluorooctanoic Acid (PFOA) contamination of drinking water resources of Arnsberg, namely the rivers M\"ohne and Ruhr, is assumed to have been comparably high from about 2004 until about 2007 \citep{Sku06, Hoe08}. PFOA is readily absorbed after ingestion, is not metabolized in the human body and has a biological half-life of 2 to 4 years. Food and water consumption are the major sources of human background exposure; in case of distinctly increased PFOA concentrations in drinking water, tap water consumption accounts for the major part of exposure \citep{Ves09}. Therefore, we include estimated concentrations in the Arnsberg drinking water as an additional covariate to the linear predictors. These concentrations are supposed to be an appropriate surrogate marker of the external exposure to PFOA, at least as a local average (cf., \citealp{Rat20} and references therein); however, the individual \textit{in}ternal exposure (PFOA concentration in blood) may not be well represented by these data without information on individual water consumption, time of residence and additional biometric data. 

Daily estimations are modeled on the spatial level of water supply stations and, afterwards, of so-called water supply areas \citep{Rat20}. They are spatio-temporally assigned to the perinatal data by postal code and day of birth, where a postal area's PFOA concentration is averaged from the respective overlapping water supply areas, weighted by population density data from the German national census of 2011 \citep{Zensus2011}. For the town of Arnsberg, we obtain the lowest PFOA concentration in the eastern part of the town (supplied by a Ruhr-dependent station upstream of the M\"ohne mouth). In the western part (supplied from M\"ohne and Ruhr, but also from stations using non-contaminated groundwater), the average concentration in 2006 is about four times higher. Finally, in the central part (solely supplied from the M\"ohne), it is almost twice as high again.

We find no significant linear effect of this estimated concentration. We also experimentally include a simplified representation of this temporally structured covariate to the models, as a first exploration of possible temporal effects despite the roughness of information: A dummy variable states whether an observation is from the years of 2004 until 2007 and, thus, from within the period of presumed very high PFOA concentrations in drinking water, but with no significant effect resulting.

\section{Discussion and Perspectives}\label{sec:disc}

\subsubsection*{Data Quality}

The secondary data from the perinatal registry have not originally been collected to be scientific material, but for quality assurance. As such, they are nonetheless very informative with regard to procedures in obstetric health care, like the birth mode (Caesarian section, induction), which turns out as an important covariate. On the other hand, measurement accuracy varies (e.g., one hospital measures birth weight (BW) accurate to 1~\si{\gram}, another to 10~\si{\gram}). This holds also for gestational age (GA), where the unit of days found in our data, instead of weeks, is not common and varying accuracy of partially rounded data is possible. In general, GA data are subject to uncertainty of reporting, measurement, clinical estimation and documentation (e.g., \citealp{Pol09}), although we have carefully checked ours for plausibility. Maternal smoking is self-reported and perhaps biased towards a socially desirable answer; nonetheless, these data are accurate enough such that an effect of smoking in line with other studies from the literature (see below) is detected despite the remaining noise.

\subsubsection*{Gestational Age and Dependence Structure}

There are strong effects of all three polynomial terms of GA in the univariate model and the increasing trend of the mean BW along GA decreases again towards the end. This phenomenon is also reported in other studies (e.g., \citealp{Skj00}) and could be an effect of medical decisions to deliver fetuses with high weights rather early by induction or sectio and to avoid such treatments for a longer time when fetal weight is low.

Against this background and given that Figure~\ref{fig:resp} hints at two tails in the data (in the regions of both high and low GA), 
our estimation of tail dependence is very sensitive to GA observation. Any data inaccuracies, which are generally possible for GA (e.g., \citealp{Pol09}), have an impact on regression models.

\subsubsection*{Model Comparison and Evaluation}

\begin{figure}[hp]
\caption{Bivariate density of birth weight and gestational age, as predicted from copula model with posterior mean parameters plugged in, conditional on certain selected exemplary covariate values (the others are fixed to: maternal height: \SI{170}{\centi\metre}, maternal BMI: \SI{20}{\kilo\gram\per\square\meter}, maternal gain of weight: \SI{10}{\kilo\gram}; and all others set to `no' or  0, respectively).}\label{fig:contour}
	\begin{center}
		\includegraphics[width=.98\textwidth]{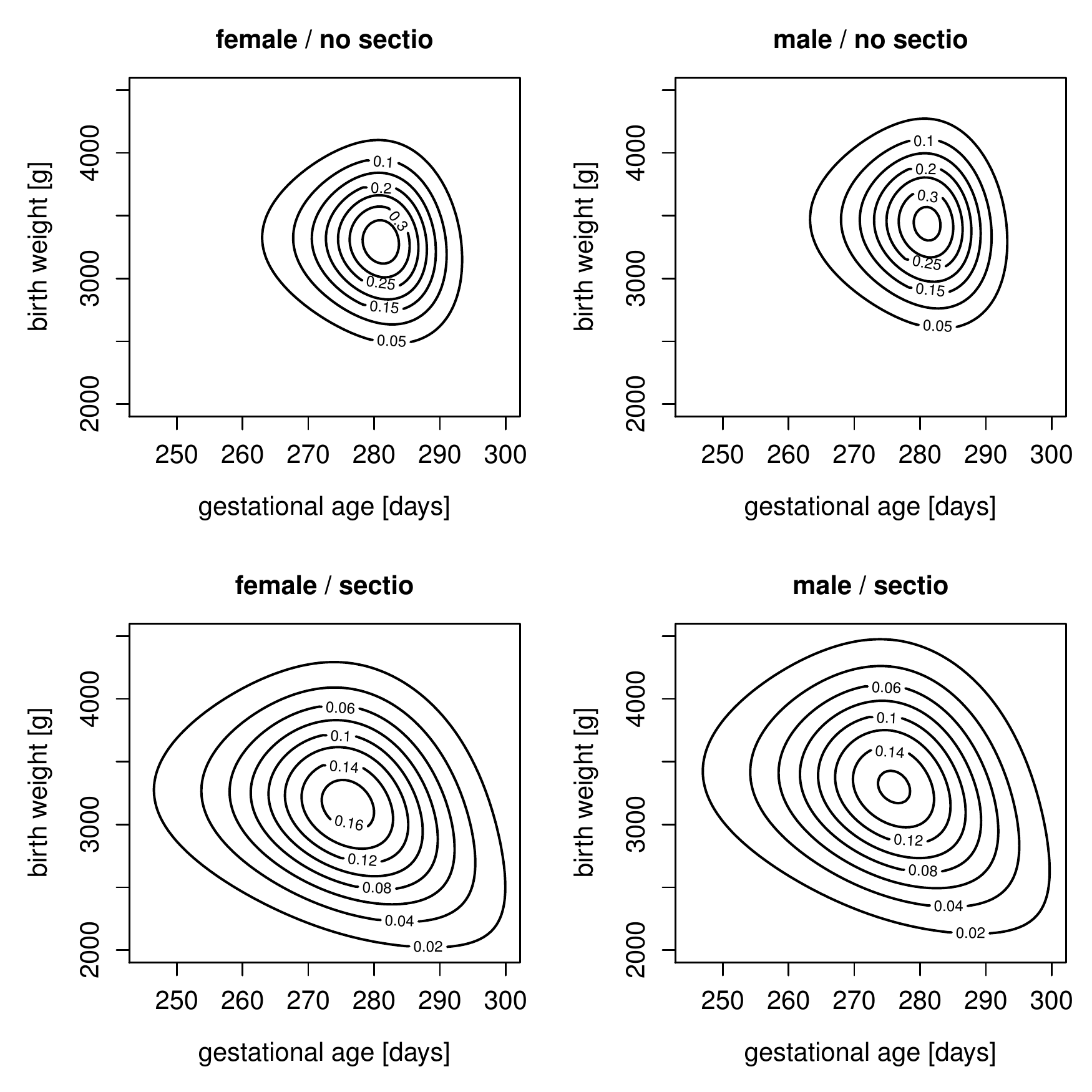} 
	\end{center}
\end{figure}

An important benefit of the distributional copula regression model are visible differences between groups, with respect to both scale and dependence: Figure~\ref{fig:contour} shows examples of predictions, distinguished by sex and sectio. It becomes apparent, that the variability and structure of the response data is deeper explained, when influences of covariates on more parameters than only the means are allowed -- unlike in a standard regression model (Section~\ref{sec:comp}).

Considering both models together, we obtain conclusions that go beyond effects of covariates on BW. A striking example are the possible relationships between BW, GA and the Cesarean section covariate (cf., Tables~\ref{tab:coeff} and \ref{tab:linpred}): The latter has a significant influence on BW according to the copula model, where GA is separately estimated, while this does not hold for the standard model, where GA is present as a significant covariate. The sectio covariate also significantly influences the parameters of the Dagum distribution of GA in the copula model as well as the copula parameter (Section~\ref{sec:eval}). According to these results, the influence of the sectio covariate is in fact manifold (cf., e.g., \citealp{Sto04}), but this can only be discovered using the bivariate model, which provides more extensive conclusions in this respect. In the standard model, the importance of the sectio covariate disappears; it is presumably predominated and in parts mediated by GA. Similar considerations hold for the induction covariate.

As a different example, both models agree with respect to the significant effect of smoking on BW (Tables~\ref{tab:coeff} and \ref{tab:linpred}). There is also an effect on GA found in the bivariate model (Section~\ref{sec:eval}, Table~\ref{tab:linpred}), but only with respect to one Dagum parameter and, thus, presumably less important.  So, there seems to be no mediation by GA in the standard model. The influence of smoking on both BW and GA as well as on the risk of pre-term birth or `small for GA' has also been found in many  studies with univariate responses (e.g., \citealp{Pol09, Kyr98,Li93}).

\subsubsection*{Integration of PFOA Exposure to Models}

A possible effect of Perfluorooctanoic Acid (PFOA) on perinatal variables is supposed to be weak and difficult to measure, compared to influences of other exposures such as maternal smoking. Furthermore, there are complex relationships and uncertainties on the path between drinking water contamination and internal exposure \citep{Kol19}. However, the mother's tap water consumption has not been recorded in the perinatal data set.   

The individual internal PFOA burden of a cohort of children from Arnsberg born about 2000 \citep{Hoe08} is known. These cohort data can be analyzed with models like ours, as far as possible with the smaller number of observations. However, these information are very useful to assess relationships between PFOA concentration in local drinking water, external and internal exposure \citep{Kol19}.

\subsubsection*{Modeling Perspectives}

The employed Dagum distribution fits well to our strongly asymmetric GA data. Other families could be possible too, but should be just as flexible and, therefore, have several parameters including shape, even when the structure of results is unfavorable for substantial interpretation. Also, other copula families as well as specific data transformations might be useful for complicated bivariate response data shapes as ours. E.g., the skewed t-copula allows for strong asymmetry and non-linearity \citep{Sun08}, but estimation and interpretation of the multiple parameters are inconvenient compared to our one-parametric representation of dependence structure.

As there remains much noise after either model fit, more complex generalized additive models, especially using splines, could be considered where non-linear relationships are possible \citep{Fah04}. This holds for the temporal modeling of PFOA concentrations, but also for the spatial dimension in future studies when larger regions are considered. There, further information such as neighbourhood of postal codes or water supply relationships could be used.  Since lower BWs are observed in some urban regions, an according spatial dependence structure can also be included. This and other model enhancements may ease the detectability of very weak effects. 

\clearpage

\section*{Acknowledgements}

Nadja Klein gratefully acknowledges support through the Emmy Noether grant KL~3037/1-1 of the German research foundation (DFG). 

The authors thank the quality assurance office (qs-nrw) at the medical association Westphalia-Lippe, in particular Hans-Joachim~B\"ucker-Nott and Heike Jaegers, for providing access to the perinatal registry data, their kind support and our helpful discussions. 

We thank Stiftung Mercator for funding parts of our work.

\bibliography{DistrCopRegr_BirthData}

\appendix

\section{Diagnostic measures for MCMC results}

\begin{table}[p]
\begin{center}
\caption{Potential scale reduction factors based on 3 chains with 1000 samples each from the posteriors of the $\hat{\beta}_j^{(\theta)}$'s in the final copula model, per parameter $\theta \in \{\mu,\sigma^2, p,a,b, \rho\}$ and covariate.}\label{tab:psrf}
\begin{tabular}{l|cc|ccc|c}
 \textbf{Response} &  \multicolumn{2}{c|}{Birth Weight} &  \multicolumn{3}{c|}{Gestational Age}  & \textbf{Copula} \\ 
 \textbf{Family} & \multicolumn{2}{c|}{Gaussian} &  \multicolumn{3}{c|}{Dagum} & Clayton \\ 
 \textbf{Parameter} & $\mu$ & $\sigma^2$ & $p$ & $a$ & $b$ & $\rho$ \\\hline 
  intercept & 0.9995 & 1.0002 & 0.9992 & 0.9991 & 0.9992 & 0.9991 \\ 
    sex (female) & 1.0008 & 1.0012 &  &  &  &  \\ 
   prev. pregn. & 1.0010 &  &  &  & 0.9993 &  \\  
   sectio & 0.9997 & 0.9991 & 0.9992 & 0.9992 & 1.0011 & 0.9994 \\   
   induction & 1.0003 &  & 0.9990 & 0.9992 & 0.9992 &  \\  
   mat. age &  &   &   &   &   &  \\ 
  mat. height & 1.0007  &  &  &  &  &  \\ 
   mat. BMI &1.0002 & 0.9995 &  &  &  &  \\ 
   weight gain & 1.0003 &  &  & 0.9998 &  &  \\  
   smoking &  0.9995 & 0.9991 &  & 0.9992 &  &  \\ 
   single & 1.0015  &   &   &   &   &  \\ 
   employed &  &   & 1.0001  &   &   &  \\
 \end{tabular}  
\end{center}
\end{table}

\begin{table}[p]
\begin{center}
\caption{Effective sample sizes based on 1000 samples from the posteriors of the $\hat{\beta}_j^{(\theta)}$'s in the final copula model, per parameter $\theta \in \{\mu,\sigma^2, p,a,b, \rho\}$ and covariate.}\label{tab:ess}
\begin{tabular}{l|cc|ccc|c}
 \textbf{Response} &  \multicolumn{2}{c|}{Birth Weight} &  \multicolumn{3}{c|}{Gestational Age}  & \textbf{Copula} \\ 
 \textbf{Family} & \multicolumn{2}{c|}{Gaussian} &  \multicolumn{3}{c|}{Dagum} & Clayton \\ 
 \textbf{Parameter} & $\mu$ & $\sigma^2$ & $p$ & $a$ & $b$ & $\rho$ \\\hline 
  intercept & 1000 & 1000 & 1000 & 1000 & 1000 & 731 \\ 
    sex (female) & 1112 & 1000 &  &  &  &  \\
   prev. pregn. & 839 &  &  &  & 1093 &  \\  
   sectio &896 & 967 & 1000 & 1000 & 1000 & 1101 \\  
   induction & 1000 &  & 1000 & 906 & 1242 &  \\  
   mat. age &  &   &   &   &   &  \\ 
  mat. height & 1000 &  &  &  &  &  \\ 
   mat. BMI &1000 & 1000 &  &  &  &  \\ 
   weight gain & 1000 &  &  & 1000 &  &  \\  
   smoking &  1000 & 1000 &  & 1000 &  &  \\ 
   single & 1000 &   &   &   &   &  \\ 
   employed &  &   & 1000 &   &   &  \\
 \end{tabular} 
\end{center}
\end{table}

\end{document}